\documentclass[fleqn,twoside,psfig]{article}
\usepackage{espcrc2}

\usepackage{graphicx}
\usepackage{epsfig}
\usepackage{axodraw}
\newcommand{\AmS}{{\protect\the\textfont2
  A\kern-.1667em\lower.5ex\hbox{M}\kern-.125emS}}

\def\simgt{\rlap{\lower 3.5 pt\hbox{$\mathchar \sim$}}\raise 1pt \hbox {$>$}}
\def\simlt{\rlap{\lower 3.5 pt\hbox{$\mathchar \sim$}}\raise 1pt \hbox {$<$}}

\newcommand{\kslash}{k\kern-1ex /}
\newcommand{\pslash}{p\kern-1ex /}
\newcommand{\qslash}{q\kern-1ex /}
\newcommand{\lslash}{l\kern-1ex /}
\newcommand{\sslash}{s\kern-1ex /}
\newcommand{\Dslash}{{\cal D}\kern-1.5ex /}

\newcommand{\beqa}{\begin{eqnarray}}
\newcommand{\eeqa}{\end{eqnarray}}

\newcommand{\be}{\begin{equation}}
\newcommand{\ee}{\end{equation}}
\newcommand{\ben}{\begin{eqnarray}}
\newcommand{\een}{\end{eqnarray}}
\newcommand{\nn}{\nonumber}
\def\lsim{\raise0.3ex\hbox{$<$\kern-0.75em\raise-1.1ex\hbox{$\sim$}}}
\def\gsim{\raise0.3ex\hbox{$>$\kern-0.75em\raise-1.1ex\hbox{$\sim$}}}
\def\simgt{\rlap{\lower 3.5 pt\hbox{$\mathchar
\sim$}}\raise 1pt \hbox {$>$}}
\def\simlt{\rlap{\lower 3.5 pt\hbox{$\mathchar
\sim$}}\raise 1pt \hbox {$<$}}

\newcommand{\ceff}{c_{\rm eff}}
\newcommand{\npcsw}{c_{\rm SW}^{\rm NP}}
\title{
\vspace*{-2.cm}
\begin{flushright}
{\normalsize UTHEP-497}\\ 
{\normalsize UTCCS-P-9}\\
\end{flushright}
Charmed meson spectra and decay constants with
one-loop $O(a)$ improved relativistic heavy quark action
\thanks{Talk presented by Y.~Kayaba}}

\author{ CP-PACS Collaboration : 
Y. Kayaba\rlap,\address{
    Grad. School of Pure and Applied Sciences,
    University of Tsukuba, Tsukuba, Ibaraki 305-8571, Japan}
S. Aoki\rlap,$^{\rm a}$ 
M. Fukugita\rlap,\address{
    Institute for Cosmic Ray Research, University of
    Tokyo, Kashiwa, Chiba 277-8582, Japan}
K-I. Ishikawa\rlap,\address{
    Department of Physics, Hiroshima University,
    Higashi-Hiroshima, Hiroshima 739-8526, Japan}
Y. Iwasaki\rlap,$^{\rm a}$ 
K. Kanaya\rlap,$^{\rm a}$, 
T. Kaneko\rlap,\address{
     High Energy Accelerator Research Organization (KEK),
     Tsukuba, Ibaraki 305-0801, Japan}
Y. Kuramashi\rlap,$^{\rm a,}$ \address{
     Center for Computational Sciences, University of
     Tsukuba, Tsukuba, Ibaraki 305-8577, Japan}
M. Okawa\rlap,$^{\rm c}$
A. Ukawa\rlap,$^{\rm a,e}$
T. Yoshi\'e\rlap,$^{\rm a,e}$ 
}

\begin{document}
\pagestyle{empty}

\begin{abstract}
We calculate charmed meson spectra and decay constants  
in lattice QCD employing one-loop $O(a)$ improved heavy quark action and 
axial-vector currents. 
In quenched simulations at $a \sim 0.1$ fm with the plaquette
gauge action as well as a renormalization-group improved one,
it is shown that the deviation from the continuum dispersion relation 
and the violation of space-time symmetry for the
pseudoscalar meson decay constants
are substantially reduced, once the $O(a)$ improvement is applied.
Preliminary results with two flavors of dynamical quarks are
also presented.
\vspace{-0.2cm}

%\vspace{1pc}
\end{abstract}

\maketitle

\section{Introduction}

With current computational resources
it is difficult to achieve a precise determination of weak
matrix elements associated with $B$ and $D$ mesons 
because of large $m_Q a$ corrections for $b$- and $c$-quarks.
Recently a relativistic heavy quark(RHQ) action
was proposed to control these $m_Q a$ corrections
by applying the on-shell $O(a)$ improvement program
to massive quarks\cite{akt}, and
the mass-dependent improvement coefficients have been  
perturbatively determined up to one-loop level\cite{cEcB}.
The improvement of the heavy-heavy and heavy-light 
axial vector currents are also carried out up to one-loop level 
in a mass dependent way\cite{zAcA}.
In this report we perform a numerical study of the RHQ action 
in quenched QCD by investigating
the effective speed of light $\ceff$ from the meson energy and
the space-time asymmetry $R$ for the pseudoscalar decay constant. 
The deviation from $\ceff=1$ and $R=1$ measures typical systematic errors
due to finite $m_Q a$ corrections
left in the formulation, which should be incorporated in the results of 
charmed meson spectra and decay constants.
Preliminary results with two
flavors of dynamical quarks are also presented.

\section{Simulation details}
We employ the clover(CL) action with non-perturbative $\npcsw$
\cite{NPCL.ALPHA,NPCL.CPPACS} for light quarks
and the RHQ action for heavy quarks combined with the plaquette
and the Iwasaki gauge actions. 
It is worth mentioning that we replace the massless part of the one-loop
improvement coefficients $c_B$ and $c_E$ in the RHQ action with
that of the non-perturbative one as
\ben
c_{B/E}=c_{B/E}^{\rm PT}(m_Q a) -c_{B/E}^{\rm PT}(0) +\npcsw,\nn
\een
where the superscript PT represents ``perturbative'' value. 
With this replacement $O(a)$ errors are completely removed at $m_Q=0$.
We calculate charmed meson spectra using $300$ configurations 
on a $24^3 \times 48$ lattice at $a \sim 0.1$fm in quenched QCD. 
Three values of the lihgt quark mass corresponding to
$M_{PS}/M_V \sim 0.56 - 0.77$ and four values of the
heavy quark mass around the charm quark mass are adopted.
For comparison we perform parallel calculations 
with the CL action for both heavy and light quarks at
the same gauge coupling and lattice size.

\begin{figure}[t]
\centering{
\psfig{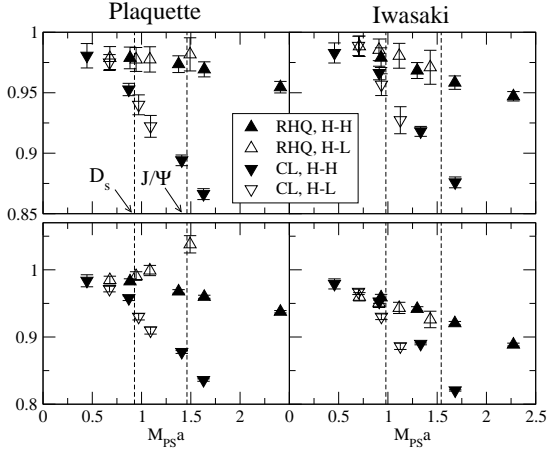}
}
\vspace{-1cm}
\caption{Effective speed of light for the PS meson(top) 
and space-time asymmetry for the PS meson decay constant(bottom) 
as a function of the meson mass in quenched QCD.}
\label{fig:Spl.AIS}
\vspace{-0.3cm}
\end{figure}

\section{Numerical study of the RHQ action}

Let us examine effectiveness of the $O(a)$ improvement for the 
RHQ action in a nonperturbative way.
We extract the effective speed of light $\ceff$ by fitting the
pesudoscalar(PS) meson energy $a E$ as a function of
spatial momentum $a p_s$: 
\[ (a E)^2 = (a M_{\rm pole})^2 +\ceff^2 (a p_s)^2, \]
where $M_{\rm pole}$ denotes the pole mass.
$\ceff$ is expected to become unity in the continuum limit.
Numerical results are plotted in Fig.~\ref{fig:Spl.AIS}(top)
as a function of the meson mass for
heavy-heavy(H-H) and heavy-light(H-L) systems. 
Around charmed meson mass the deviation from $\ceff=1$
is equal to or larger than about $10\%$ for the CL heavy quark action, 
while less than $5\%$  for the RHQ action.

We also measure the space-time asymmetry in terms of the PS meson 
decay matrix elements: 
\[ R \equiv i \frac{\langle 0|A_k^R|PS\rangle}
{\langle 0|A_4^R|PS\rangle}\cdot\frac{E}{|p_s|}, \]
which should become unity in the continuum limit.
Results are plotted in Fig.\ref{fig:Spl.AIS}(bottom) as a function of the 
meson mass.
The asymmetry reaches $10 \sim 20\%$ 
around the charmed meson mass for the CL action,
while it is less than $8\%$  for the RHQ action.
These observations allow us to conclude that the improved action
works as designed in reducing the $m_Q a$ corrections.

\begin{figure}[t]
\centering{
\psfig{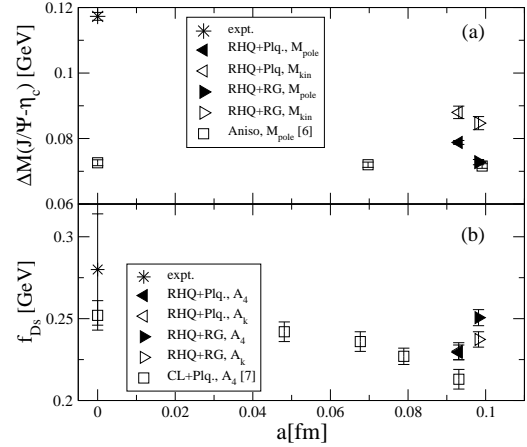}
}
\vspace{-1cm}
\caption{Charmonium S-state hyperfine splitting(top)
and $D_s$ meson decay constant(bottom) in quenched QCD
as a function of $a$ set by Sommer scale $r_0$.}
\label{fig:fDs}
\vspace{-0.3cm}
\end{figure}

%\section{Spectra and decay constants in $N_f=0$ QCD}

Figure~\ref{fig:fDs}(a) shows the charmonium S-state hyperfine
splitting as a function of lattice spacing, where we practice two
analysis: one is based on the pole mass of the charmed meson 
and another on the kinetic mass: 
\[ a M_{\rm kin}= a M_{\rm pole}/\ceff^2. \]
We find $15\%$ discrepancy, which should be included 
as a systematic error. Compared to results with
anisotropic lattice QCD\cite{Aniso}, 
our results are slightly larger at $a \sim 0.1$fm.
Both are substantially lower than the experimental value.

In Fig.~\ref{fig:fDs}(b) we plot the $D_s$ meson decay
constant which are obtained from the time and spatial
components of the axial vector current.
The difference between two definitions is less than $3\%$ for both
gauge actions. For comparison we also plot the results obtained by  
the CL heavy quark action with the plaquette gauge action\cite{fDs.ALPHA}.
Our results at $a\sim 0.1$fm are much 
closer to the value of the CL heavy quark action
in the continuum limit than at a similar lattice spacing.
This would suggest that the RHQ action indeed improves 
the scaling behavior of the decay constant.

\begin{figure}[t]
\centering{
\psfig{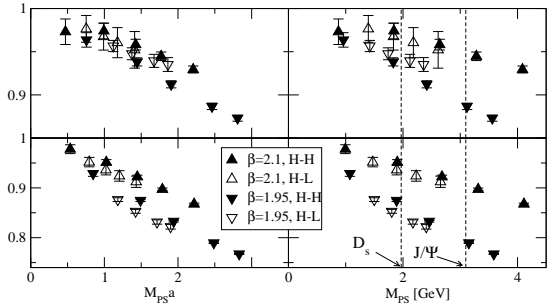}
}
\vspace{-0.8cm}
\caption{Effective speed of light for the PS meson(top) 
and space-time asymmetry for the PS meson decay constant(bottom) 
as a function of the meson mass in lattice unit(left) 
and physical unit(right) for $N_f=2$ QCD.}
\label{fig:Spl.AIS.Nf2}
\vspace{-0.3cm}
\end{figure}

\section{Preliminary results of $N_f=2$ QCD}

Encouraging results in quenched QCD urge us to apply
the calculation to the two-flavor QCD gauge
configurations\cite{Nf2Conf}. We employ the mean-field 
improved CL light quark action and 
the RHQ action for the valence quraks.
Up to now two set of lattices have been analyzed:
400/200 configurations for $16^3 \times 32$/$24^3 \times 48$ 
at $\beta=1.95/2.1$ with four sea quark masses.
Lattice spacings are roughly $0.15$fm and $0.11$fm, respectively.
We make a similar choice for the set of valence heavy 
and light quark masses as in the quenched case.

In Fig.~\ref{fig:Spl.AIS.Nf2} we show the effective speed
of light and the space-time asymmetry for the PS meson decay constant. 
As $a$ decreases, the deviation from $\ceff=1$ 
and $R=1$ is reduced at fixed $m_Q a$ and $m_Q$.
These behavior are consistent with the expected 
leading scaling violation of the RHS action given by  
$O(f_2(m_Q a)(a \Lambda_{\rm QCD})^2)$ or $O(f_1(m_Q
a) g^4 a \Lambda_{\rm QCD})$, where $f_1$ and $f_2$ 
are some unkown functions.
A little surprising is that, as seen from the left panels in 
Fig.~\ref{fig:Spl.AIS.Nf2},
$f_{1,2}$ look almost linear in terms of $m_Qa$ even beyond $m_Qa=1$. 

\begin{figure}[t]
\centering{
\psfig{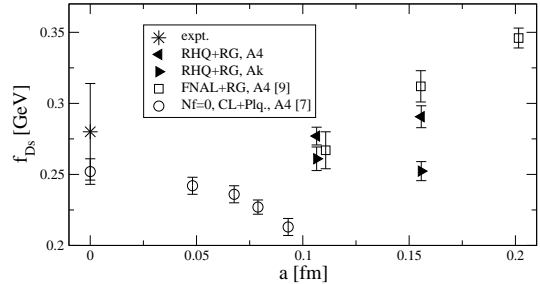}}
\vspace{-1cm}
\caption{$D_s$ meson decay constant for $N_f=2$ QCD as a
function of $a$. The scale is set by $r_0$
for our results and $m_\rho$ for FNAL method.}
\label{fig:fDs:Nf2}
\vspace{-0.3cm}
\end{figure}

Figure~\ref{fig:fDs:Nf2} illustrates the $a$ dependence  of the $D_s$ meson 
decay constant. 
While the scaling behavior from $A_4$ with the RHQ
action is only a little milder 
over the previous result obtained by the FNAL interpretation
with the CL heavy quark action\cite{fDs.FNAL},
the one from $A_k$ is much improved.
This result suggests that we will be able to make a reliable continuum 
extrapolation, employing a simultaneous fit for $f_{D_s}$ obtained 
from $A_4$ and $A_k$, once we have data on CP-PACS configurations
at $\beta =2.2$ or 1.8.

This work is supported in part by the Grants-in-Aid for
Scientific Research from the Ministry of Education,
Culture, Sports, Science and Technology.
(Nos.~
13135204, %Iwasaki
13640260, %Kanaya
14046202, %Aoki
14740173, %Kaneko
15204015, %Ukawa
15540251, %Aoki
15540279, %Okawa
%15740134, %ishizuka
16028201, %Aoki
16540228, %Yoshie
16740147, %Ishikawa
)

\end{document}